\documentclass[letterpaper,10pt,twocolumn,final,journal,oneside]{IEEEtran}
\usepackage{multirow}
\usepackage{graphicx}
\usepackage{acronym}
\usepackage{amsmath}
\usepackage{eurosym}
\usepackage{siunitx}
\usepackage{stackengine}
\usepackage{tikz}
\usepackage{textcomp}
\usepackage{hyperref}
\usepackage{ccicons}
\usepackage{tensor}
\usepackage{balance}

\acrodef{lu}[LU]{load unit}
\acrodef{pev}[PEV]{plug-in electric vehicle}
\acrodef{tcl}[TCL]{temperature controlled load}
\acrodef{abp}[ABP]{appliance based on phases}
\acrodef{bess}[BESS]{battery energy storage system}
\acrodef{res}[RES]{renewable energy source}
\acrodef{agg}[AGT]{aggregator}
\acrodef{dap}[DAP]{day-ahead planning}
\acrodef{tso}[TSO]{transmission system operator}
\acrodef{dso}[DSO]{distribution system operator}
\acrodef{pv}[PV]{photo-voltaic}
\acrodef{upd}[UPD]{user-programmable device}
\acrodef{ncd}[NCD]{non-controllable device}
\acrodef{soc}[SoC]{state-of-charge}
\acrodef{dr}[DR]{demand response}
\acrodef{tou}[TOU]{time of use}
\acrodef{der}[DER]{distributed energy resource}
\acrodef{rtp}[RTP]{real-time pricing}
\acrodef{cpp}[CPP]{critical peak pricing}

\def \soc {{\rm SoC}}

\DeclareSIUnit{\kWh}{kWh}

\newcommand\copyrighttext{%
  \centering\ccbyncnd\\ \copyright~2020 the authors. This work has been published under the\\
  \href{http://creativecommons.org/licenses/by-nc-nd/4.0/}{Creative Commons Attribution-NonCommercial-NoDerivatives 4.0 International License.}\\
  21st IFAC World Congress, vol.~53, no.~2, pp.~12213--12220 DOI: \href{https://doi.org/10.1016/j.ifacol.2020.12.1098}{10.1016/j.ifacol.2020.12.1098}.}
\newcommand\copyrightnotice{%
\begin{tikzpicture}[remember picture,overlay]
\node[anchor=north,yshift=0pt] at (current page.north) {\setlength{\fboxrule}{0pt}\fbox{\parbox{\dimexpr\textwidth-\fboxsep-\fboxrule\relax}{\copyrighttext}}};
\end{tikzpicture}%
}

\begin{document}

\title{An Optimization Problem for Day-Ahead Planning of Electrical Energy Aggregators}

\author{Francesco~Conte, Matteo~Saviozzi, Samuele~Grillo%
\thanks{F. Conte and M. Saviozzi are with  Universit\`a degli Studi di Genova, DITEN, via all'Opera Pia,11/A, 16145 Genova GE, Italy (e-mail: fr.conte@unige.it, matteo.saviozzi@unige.it).}%
\thanks{S. Grillo, is with the Dipartimento di Elettronica, Informazione e Bioingegneria, %
Politecnico di Milano, piazza Leonardo da Vinci, 32, I-20133 Milano, Italy %
(e-mail: samuele.grillo@polimi.it).}%
}

\IEEEaftertitletext{\copyrightnotice\vspace{1.1\baselineskip}}
\maketitle

\begin{abstract}                
The widespread diffusion of distributed energy resources, especially those based on renewable energy, and energy storage devices has deeply modified power systems. As a consequence, demand response, the ability of customers to respond to regulating signals, has moved from large high-voltage and medium-voltage end-users to small, low-voltage, customers. In order to be effective, the participation to demand response of such small players must be gathered by aggregators. The role and the business models of these new entities have been studied in literature from a variety of viewpoints. Demand response can be clearly applied by sending a dedicated price signal to customers, but this methodology cannot obtain a diverse, punctual, predictable, and reliable response. These characteristics can be achieved by directly controlling the loads units. This approach involves communication problems and technological readiness. This paper proposes a fully decentralized mixed integer linear programming approach for demand response. In this framework, each load unit performs an optimization, subject to technical and user-based constraints, and gives to the aggregator a desired profile along with a reserve, which is guaranteed to comply with the constraints. In this way, the aggregator can trade the reserve coming from several load units, being the only interface to the market. Upon request, then, the aggregator communicates to the load units the modifications to their desired profiles without either knowing or caring how this modification would be accomplished. The effectiveness is simulated on 200 realistic load units.
\end{abstract}

\begin{IEEEkeywords}
demand response, load aggregator, control of renewable energy resources, intelligent control of power systems, optimal operation and control of power systems, smart grids.
\end{IEEEkeywords}

\section{Introduction}

\IEEEPARstart{P}{ower} systems are  experiencing an important modification both in their structure and, consequently, in the way they are operated. One of the results of these changes is the fact that customers (passive and irresponsive in the previous paradigm) have become active and potentially effective resources for stability and control of power systems~\cite{Hu:2017}.

The way through which this could be accomplished is called \ac{dr}. Traditionally, \ac{dr} has been applied and implemented in the industrial sector~\cite{Ayon:2017}. If \ac{dr} is to be applied to residential customers, an additional player between \ac{tso} and producers has to be introduced. This new player is the \ac{agg}, whose role is to gather a (potentially large) number of customers, collect their availability and trade it into dedicated service markets~\cite{Zhou:2016}.

There is a vast literature on \acp{agg} and, specifically, on their role in power systems and markets. Their effectiveness and, related to this, the need and convenience to introduce them, is questionable. In fact, \ac{dr} can naturally be obtained through adequate price schemes, such as \ac{rtp}, \ac{tou}, and \ac{cpp}~\cite{Li:2016,Alizadeh:2012}. However, the usage of such a broadcast signal (\textit{i.e.}, the price of energy) can reduce the ability to obtain a predictable and reliable response~\cite{Alizadeh:2012}. Moreover, these approaches tend to limit the service provision to a load reduction~\cite{Chen:2012}.

In~\cite{Gkatzikis:2013} the role of \acp{agg} is explored. In this work a \ac{dr} market scheme is described, along with the interactions among \acp{dso}, \acp{agg}, and end users. The main objective is to define an overall framework in which all these interactions prove their effectiveness. In~\cite{Li:2016,Zhu:2019} a market-based framework for residential \acp{tcl} is developed. Thus, controllable loads are supposed to react to energy prices. In~\cite{Parvania:2013} the business models of \acp{agg} are analyzed and an optimization of the participation of \acp{agg} to \ac{dr} markets is presented. The focus of the work is on the day-ahead operation of \acp{agg} and the maximization of their revenues. The same problem is analyzed in~\cite{Chen:2012}, where authors apply a Cournot-based game model, and in~\cite{CorreaFlorez:2018}, where battery degradation estimation is included in the optimization problem.

On the other side, in~\cite{Alizadeh:2012} a control architecture is proposed, under the assumption that loads can be directly controlled. Also in~\cite{Ayon:2017} loads are supposed to be controlled by \acp{agg}. Authors introduce the concept of the hourly bounds that each load can provide to \ac{dr}. However, also in this work the optimization is performed at \ac{agg} level. Moreover, since a more detailed knowledge would be unlikely, an hourly demand flexibility ratio is defined based on maximum and minimum powers for each customer without any correlation with past and future behavior.

In~\cite{Iria:2019} both the \ac{agg}-level optimization and the real-time control of resources is proposed. The aim of the work is to describe a hierarchical methodology (based on model predictive control) for \acp{agg} to optimize the provision of different market products. Also in this case, resources are supposed to be directly controllable. A similar objective is pursued in~\cite{Henriquez:2018}, where \acp{agg} manage a portfolio of \ac{dr} products in day-ahead markets.

The present paper proposes a fully-decentralized optimization framework of \acp{agg} for \ac{dr}. Each \ac{lu}, equipped with different appliances (\textit{e.g.}, appliances based on phases, \acp{bess}, \acp{pev}, \acp{tcl}, etc.), performs mixed integer linear programming optimization problem and provides to the \ac{agg} a desired load profile, along with a positive and a negative reserve, for the following day. The peculiarity of this approach is that the reserve, apart from being guaranteed, complies with the multiple technical and behavioral (\textit{i.e.}, those imposed by users) constraints. The main contributions of the paper are: i) the decentralized nature of the approach; ii) the intrinsic feasibility of the reserves of each \ac{lu}; iii) the ability of the approach to compensate through the reserve both forecast and modeling errors; and iv) the detailed description of the models of the equipment installed in \acp{lu}. The paper is organized as follows. In Section~\ref{sec:aggregate_dap_and_day_operation} \ac{dap} strategy and intra-day operation rules for the \ac{agg} are presented; Section~\ref{sec:lu_models} is devoted to the description of the models and the constraints of the equipment in each \ac{lu}; Section~\ref{sec:lu_dap} presents the \ac{dap} optimization problem for the single \acp{lu}; in Section~\ref{sec:simulation_results} the results from the simulations are shown; finally, in Section~\ref{sec:conclusions} conclusions are drawn.

\section{Aggregate DAP and day operation} \label{sec:aggregate_dap_and_day_operation}
\begin{figure}[t]
	\centering
	\includegraphics[width=0.76\columnwidth]{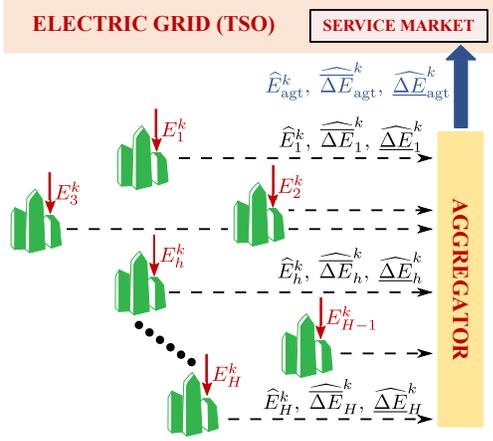}
	\caption{Aggregate day-ahead planning (DAP) scheme.}\label{fig:dap}
\end{figure}

As shown in Fig. \ref{fig:dap}, we consider an aggregate composed by $H$ \acp{lu} managed by an \ac{agg}. For each day, the day before, a \ac{dap} is computed; then, during the day, the plan is realized by following suitably established rules.

In the next, we illustrate first the \ac{dap} strategy and then the rules adopted during the intra-day operations.

\subsection{Day-Ahead Planning}
Let us assume that the day-time is partitioned in $T$ sampling intervals lasting $\Delta t$ [\si{\hour}] (\textit{e.g.}, \SI{15}{\minute} = \SI{0.25}{\hour} ). At time step $k$ ($k=0,1\ldots,T-1$), the $h$th ($h=1,2,\ldots,H$) \ac{lu} exchanges with the grid the energy $E_h^k$ [\si{\kilo  Wh}]. If $E_h^k\geq 0$ the $h$th \ac{lu} is importing energy, whereas, if $E_h^k< 0$ it is exporting energy. Notice that the energy export is a modality which can be possible or not depending on the case.

All days, each \ac{lu} locally solves a \ac{dap} optimization problem (LU-DAP) returning, for any time step $k$, the \textit{optimized energy exchange} $\widehat{E}_h^{k}$, and the \textit{optimized positive and negative energy reserves} $\widehat{\overline{\Delta E}}_h^k\geq0$ and $\widehat{\underline{\Delta E}}_h^k\leq0$ [\si{\kWh}]. These two last quantities are computed as the maximal positive and negative deviations from $\widehat{E}_h^{k}$ that the $h$th \ac{lu} declares to be able to guarantee. Notice that the load convention is used; therefore, a positive reserve corresponds to an increase of load and \textit{viceversa}.

The $h$th \ac{lu} pays for the imported energy the price $c_{{\rm imp},h}^k$ [\EUR{}/\si{\kWh}] and is (possibly) paid for the exported energy with the price $c_{{\rm exp},h}^k$ [\EUR{}/\si{\kWh}]. These prices are negotiated by the single \acp{lu} directly with the energy provider. At the same time, the \ac{agg} participates to the service market~\cite{Iria:2019}. Specifically, it receives from the \acp{lu} their optimized energy profiles and energy reserves and declares to the market the \textit{aggregated base energy profile}, defined as
\begin{equation}\label{eq:Pagg}
   \widehat{E}_{\rm agt}^k = \sum_{h=0}^{T-1} \widehat{E}_h^k,
\end{equation}
and offers the \textit{aggregated positive and negative reserves}:
\begin{equation}\label{eq:Eagg}
    \widehat{\overline{\Delta E}}^k_{\rm agt}= \sum_{h=1}^H  \widehat{\overline{\Delta E}}^k_{h}, \quad \widehat{\underline{\Delta E}}^k_{\rm agt}= \sum_{h=1}^H  \widehat{\underline{\Delta E}}^k_{h}.
\end{equation}

The \ac{agg} is paid for the aggregated reserve with the price $\tensor*[^k]{c}{*_{\rm flx}^{\rm agt}}$ [\EUR{}/\si{\kWh}], and pays the single \acp{lu} for their reserves $\tensor*[^k]{c}{*_{\rm flx}}$[\EUR{}/\si{\kWh}]. These prices are assumed to be the same for positive and negative reserves, and for each \acp{lu}. This assumption is reasonable and coherent with the existing examples of service markets~\cite{Iria:2019}.

The resulting programmed income of the \ac{agg} is
\begin{equation}
    I_{\rm agt} = \sum_{k=0}^T \left(\tensor*[^k]{c}{*_{\rm flx}^{\rm agt}}- \tensor*[^k]{c}{*_{\rm flx}} \right)\left(\overline{\Delta E}^k_{\rm agt} - \underline{\Delta E}^k_{\rm agt}\right).
\end{equation}
It is worth remarking that in this paper a fully distributed solution is proposed. Indeed, the \ac{dap} optimization is operated only locally by the single \acp{lu}. Given a value of the price $\tensor*[^k]{c}{*_{\rm flx}}$ paid by the \ac{agg} to the \acp{lu} for the energy reserves, $I_{\rm agt}$ results to be maximized by the solutions of the distributed \acp{dap}.
However, the \ac{agg} could influence the decisions of the single \acp{lu} by modifying the price $\tensor*[^k]{c}{*_{\rm flx}}$, leading them to offer a larger or a lower amount of energy reserve. This will make the problem nonlinear and a centralized or partially distributed solution should be adopted. However, this is beyond the scope of the paper, which is mainly focused on studying how a \ac{lu} can guarantee an energy reserve. Future works will regard to this kind of developments.

\subsection{Intra-Day operation}
During the considered day, the \ac{tso} can require an energy profile (\ac{dr} signal) of the form:
\begin{equation}
    E_{\rm agt, ref}^k = \widehat{E}_{\rm agt}^k + \Delta {E}_{\rm agt, ref}^k,
\end{equation}
where the variation $\Delta {E}_{\rm agt,ref}^k$ is such that
\begin{equation}
\widehat{\underline{\Delta {E}}}_{\rm agt}^k \leq    \Delta {E}_{\rm agt,ref}^k \leq \widehat{\overline{\Delta {E}}}_{\rm agt}^k.
\end{equation}

To provide the required energy reserve, the \ac{agg} distributes an energy variation reference signal to the $H$ \acp{lu} of the form:
\begin{equation}\label{eq:Eref}
    {\Delta E}_{{\rm ref},h}^k = \overline{\gamma}_h^k\max(0,\Delta {E}_{\rm agt,ref}^k) + \underline{\gamma}_h^k\min(0,\Delta {E}_{\rm agt,ref}^k)
\end{equation}
where $h=1,2,\ldots,H$ and
\begin{equation}
    \overline{\gamma}_h^k = \frac{\widehat{\overline{\Delta E}}_h^k}{\widehat{\overline{\Delta {E}}}_{\rm agt}^k}, \quad \underline{\gamma}_h^k = \frac{\widehat{\underline{\Delta E}}_h^k}{\widehat{\underline{\Delta {E}}}_{\rm agt}^k}.
\end{equation}

In this way, the actual contribution of the $h$th \ac{lu} is proportional to the percentage of the positive and negative energy reserves that it has declared to be able to guarantee.

\section{Load Unit Devices Models} \label{sec:lu_models}
Let us focus on the single $h$th \ac{lu}. The general assumption is that it is equipped with: a) \textit{smart appliances}, able to communicate and make decisions; b) \textit{smart plugs}, able to acquire, transmit measurements, and allow a remote on/off control; c) \textit{user interfaces}, that allow users to configure the desired behaviour of the controlled devices; d) \textit{a local communication network and a gateway}, which are necessary to control and monitor all the different devices and to have an interface with the Internet and with the \ac{agg}; and e) a \textit{local management system} which has computational capabilities.

The devices supposed to be installed on the $h$th \ac{lu} are:
\begin{enumerate}
    \item[1)] $N_{\rm abp}$ \textbf{\ac{abp}}, such as dish washers, laundries, industrial process appliances;
    \item[2)] $N_{\rm pev}$ \textbf{\acp{pev}} for commercial and/or personal use;
    \item[3)] $N_{\rm bess}$ \textbf{\acp{bess}}, suitably installed to obtain flexibility;
    \item[4)] $N_{\rm tcl}$ \textbf{\acp{tcl}}, such as fridges, air heating/cooling systems, water heaters;
    \item[5)] $N_{\rm res}$ \textbf{\acp{res}}, such as \ac{pv} generators;
    \item[6)] $N_{\rm upd}$ \textbf{\acp{upd}}, such as lights for industrial and/or commercial use;
    \item[7)] \textbf{\ac{ncd}},  such as TVs, PCs, domestic lights, domestic ovens, hair dryers.
\end{enumerate}

In the following, each of these devices type is described and modeled thorough a set of mixed-integer constraints, that are finally included in the \ac{lu}-\ac{dap} optimization problem. The objective of \ac{lu}-\ac{dap}, which will be formalized after in the paper by a cost function, is to minimize the day economical cost of the \ac{lu}, taking into account the possible provision of energy reserves pad by the \ac{agg}. Because of their characteristics, \acp{abp} and \acp{pev} cannot be used to directly provide energy reserves. However, they can offer a time flexibility, which can be exploited to optimize the \ac{lu} plan. Differently, \acp{bess} and \acp{tcl} can directly offer energy reserves, using the storage of electric and thermal energies that they respectively manage. Therefore, the models illustrated in the following for these two types of devices are designed in order to quantify the potential energy reserves they can provide.

It is worth remarking that the proposed models are developments of the ones introduced in~\cite{Baccino:2014} for smart houses. In all models the day time is partitioned as done for the aggregate in Section~\ref{sec:aggregate_dap_and_day_operation} with the sampling time $\Delta t$ and time steps are indicated with $k=0,1,\ldots,T-1$.

\subsection{Appliances based on phases model}\label{ssec:abp_model}
Examples of \acp{abp} are dish washers and washing machines in houses, hotels and restaurants, or process machines in factories. These appliances are characterized by a service to be provided to the user and a corresponding set of ordered working phases necessary to realize such a service. The peculiarity of a working phase is that it cannot be interrupted. Whereas, depending on the case, there is a maximal possible delay between the execution of two consecutive phases.
The user is usually interested in obtaining the service up to a certain time. Therefore, the idea is that, the day before, the user indicates the required service and the time preferences for the day-ahead. In the following, we will show how this indication can be included in the optimization model.

Let $j=1,2,\ldots,n_i$ be the index that identify the $n_i$ working phase of the $i$th \ac{abp}, $i=1,2,\ldots,N_{\rm abp}$. The following variables defined for any time steps $k$:
$p_{ij}^k$ [\si{\kilo\watt}], power absorbed by $j$th phase; $x_{ij}^k$, binary variable which identifies the activation state of $j$th phase; $s_{ij}^k$, binary variable equal to 1 if $j$th phase has been already executed at time step $k$; $t_{ij}^k$ binary variable equal to 1 if, at time step $k$, the \ac{abp} is passing from $(j-1)$th to $j$th phase


The behaviour of the $i$th \ac{abp} can be therefore described by the operation constraints listed in the following:
\begin{equation}\label{eq:abp_c1}
  \Delta t \sum_{k=0}^{T-1}p_{ij}^k = E_{ij}\qquad \forall i,j,
\end{equation}
which sets the total energy used by the $j$th phase to the parameter $E_{ij}$ [\si{\kWh}];
\begin{equation}\label{eq:abp_c2}
  \sum_{k=1}^{T}x_{ij}^k = \overline{T}_{ij}\qquad \forall i,j,
\end{equation}
which sets the total execution time of the $j$th phase to the parameter $\overline{T}_{ij}$ [\# of time steps];
\begin{equation}\label{eq:abp_c3}
  x_{ij}^k {P}^{\rm min}_{ij} \leq p_{ij}^k \leq x_{ij}^k {P}^{\rm max}_{ij}\qquad \forall i,j,k,
\end{equation}
which imposes the power limits for the $j$th phase between the maximum and minimum values $P^{\rm max}_{ij}$ and $P^{\rm min}_{ij}$ [\si{\kilo\watt}];
\begin{align}
         x_{ij}^k + s_{ij}^k\leq 1&\qquad  \forall i,j,k \label{eq:abp_c4}\\
         x_{ij}^{k-1} - x_{ij}^k - s_{ij}^k\leq 0 &\qquad \forall i,j, k=1,2,\dotsc ,T-1 \label{eq:abp_c5}\\
         s_{ij}^{k-1} - s_{ij}^k\leq 1&\qquad  \forall i,j, k=1,2,\dotsc ,T-1  \label{eq:abp_c6}\\
         x_{ij}^k - s_{i(j-1)}^k \leq 0 &\qquad \forall i,k, j=2,3,\dotsc , n_i, \label{eq:abp_c7}
\end{align}
which set the order of the phases and avoid their interruption;
\begin{align}
         t_{ij}^k = s_{i(j-1)}^k - \left ( x_{ij}^k + s_{ij}^k\right )& \qquad \forall i,j,k \label{eq:abp_c8}\\
         \sum_{k=1}^{T}t_{ij}^k \leq D_{ij}& \qquad \forall i, j=2,3,\dotsc ,n_i \label{eq:abp_c9}
\end{align}
which set to $D_{ij}$ [\# of time steps] the maximal delay between the $(j-1)$th and the $j$th phase;
\begin{equation}
  x_{ij}^k \leq {\rm UP}_{{\rm abp},i}^k \qquad \forall i,j,k \label{eq:abp_c10}
\end{equation}
which assures that the \ac{abp} is activated when desired by the user. ${\rm UP}_{{\rm abp},i}^k$ is a binary parameter that defines (when equal to 1) if the \ac{abp} can run at time step $k$, based on the user preferences. For example, assume that the required service must be completed up to a certain hour of the day, corresponding to the time step $k^*$. Then, ${\rm UP}_{{\rm abp},i}^k$=1 for $k=0,1,\ldots,k^*$ and ${\rm UP}_{{\rm abp},i}^k$=0 for the remaining entries.
\subsection{PEVs model}
Let us consider the $i$th \ac{pev}, $i=1,2,\ldots,N_{\rm pev}$. To model this type of device, two variables, defined for any time step $k$, are required: $p_{i}^k$~[\si{\kilo\watt}], absorbed power; $x_{i}^k$, binary variable which identifies the \ac{pev} recharge activation state.

The operational constraints are:
\begin{equation}\label{eq:pev_c1}
  \Delta t \sum_{k=0}^{T-1}\eta_i p_{ij}^{k} = \Delta {\rm SoC}_{i}^{\rm p}E_{{\rm p},i}^{\rm nom} \quad \forall i,
\end{equation}
which sets the final recharge amount to $\Delta {\rm SoC}_{i}^{\rm p}\geq 0$~[p.u.], and where $E_{{\rm p},i}^{\rm nom}$~[\si{\kWh}] is the nominal energy of the battery and $\eta_i\leq 1$ is the recharge efficiency;
\begin{equation}\label{eq:pev_c2}
  0 \leq p_{i}^k \leq x_{i}^k P^{\rm max}_{i}\qquad \forall i,k,
\end{equation}
which limits the recharge power under the rated value $P^{\rm max}_i$~[\si{\kilo\watt}];
\begin{equation}\label{eq:pev_c3}
  x_{i}^k \leq {\rm UP}_{{\rm pev},i}^k \qquad \forall i,k,
\end{equation}
which assures that the \ac{pev} is recharged according to the user preferences, represented by the binary parameters ${\rm UP}_{{\rm pev},i}^k$, defined as explained for the \acp{abp}.

\subsection{BESSs model}\label{ssec:bess_model}
Let us consider the $i$th \ac{bess}, $i=1,2,\ldots,N_{\rm bess}$. The basic variables, defined for any time step $k$ are: ${\rm{SoC}}_i^k$~[p.u.], battery \ac{soc}; $p_{{\rm ch},i}^k$~[\si{\kilo\watt}], charging power; $p_{{\rm dsc},i}^k$~[\si{\kilo\watt}], discharging power; $x_{{\rm ch},i}^k, \ x_{{\rm dsc},i}^k$, binary variables which identifies charging and discharging activation states, respectively.

The \ac{soc} time evolution is modeled by
\begin{equation}\label{eq:soc_dyn}
    \soc_i^k = \soc^{k-1}_i +\frac{\Delta t}{E_{{\rm b},i}^{\rm nom}} \left( \eta_i^{\rm ch} {p}_{{\rm ch},i}^{k-1} +\eta_i^{\rm dsc}  {p}_{{\rm dsc},i}^{k-1} \right)
\end{equation}
for $k=1,\ldots,T$, given the initial condition $\soc_i^0$. In \eqref{eq:soc_dyn}: $\eta^{\rm ch}_i\leq 1$ and $\eta^{\rm dsc}_i\geq 1$ are the charging and discharging efficiencies, respectively, $E_{{\rm b},i}^{\rm nom}$ [\si{\kWh}] in the nominal energy of the battery. Model \eqref{eq:soc_dyn} is completed by the following constraints:
 \begin{align}
         0\leq p_{{\rm ch},i}^{k}\leq x_{{\rm ch},i}^{k}{P}^{\rm max}_{{\rm ch},i} \qquad \forall i,k, \label{eq:bess_c1}\\
         x_{{\rm dsc},i}^{k} P^{\rm min}_{{\rm dsc},i} \leq p_{{\rm dsc},i}^{k} \leq 0 \qquad\forall i,k, \label{eq:bess_c2} \\
         x_{{\rm ch},i}^{k} + x_{{\rm dsc},i}^{k} \leq 1 \qquad \forall i,k,   \label{eq:bess_c3}
\end{align}
which limit the power exchange of the \ac{bess} between the maximum recharge power $P^{\rm max}_{{\rm ch},i}\geq 0$ and the maximum (in absolute value sense) discharge power $P^{\rm min}_{{\rm dsc},i} \leq 0$, and allow the activation of only one task (charge or discharge) at any time.

In order to quantify the potential energy reserve that can be provided during the day by the \ac{bess}, the maximal positive and negative variations of charging and discharging powers are introduced: $\overline{\Delta p}_{{\rm ch},i}^k, \overline{\Delta p}_{{\rm dsc},i}^k \geq 0$ and $\underline{\Delta p}_{{\rm ch},i}^k, \underline{\Delta p}_{{\rm dsc},i}^k \leq 0$  [\si{\kilo\watt}]. These variations will imply two maximal deviations from the the base battery power exchange profile $p_{{\rm bess},i}^k=p_{{\rm ch},i}^k+p_{{\rm dsc},i}^k$:
\begin{equation} \label{eq:over_delta_pbess}
    \overline{\Delta p}_{{\rm bess},i}^k = \overline{\Delta p}_{{\rm ch},i}^k + \overline{\Delta p}_{{\rm dsc},i}^k \quad \forall i,k,
\end{equation}
\begin{equation}\label{eq:under_delta_pbess}
    \underline{\Delta p}_{{\rm bess},i}^k = \underline{\Delta p}_{{\rm ch},i}^k + \underline{\Delta p}_{{\rm dsc},i}^k \quad \forall i,k,
\end{equation}
which can therefore be used to provide energy reserves. Given the maximal variation variables, it is also possible to compute the time evolution of the \ac{soc} over-bound $\overline{\soc}^k_i$ and under-bound $\underline{\soc}^k_i$ as it follows:
\begin{equation}\label{eq:over_soc_dyn}
    \begin{aligned}
    \overline{\soc}_i^k = \overline{\soc}_i^{k-1} &+ \frac{\Delta t}{E_{{\rm b},i}^{\rm nom}}\eta_i^{\rm ch} \left(p_{{\rm ch},i}^{k-1} + \overline{\Delta p}_{{\rm ch},i}^{k-1}\right) \\
    &+  \frac{\Delta t}{E_{{\rm b},i}^{\rm nom}}\eta_i^{\rm dsc} \left(p_{{\rm dsc},i}^{k-1} + \overline{\Delta p}_{{\rm dsc},i}^{k-1}\right)
    \end{aligned}
\end{equation}
\begin{equation}\label{eq:under_soc_dyn}
    \begin{aligned}
    \underline{\soc}_i^k = \underline{\soc}_i^{k-1} &+ \frac{\Delta t}{E_{{\rm b},i}^{\rm nom}}\eta_i^{\rm ch} \left(p_{{\rm ch},i}^{k-1} + \underline{\Delta p}_{{\rm ch},i}^{k-1}\right) \\
    &+  \frac{\Delta t}{E_{{\rm b},i}^{\rm nom}}\eta_i^{\rm dsc} \left(p_{{\rm dsc},i}^{k-1} + \underline{\Delta p}_{{\rm dsc},i}^{k-1}\right)
    \end{aligned}
\end{equation}
$\forall i,k=1,2,\ldots,T$, with inital conditions $\overline{\soc}_i^0$ and $\underline{\soc}_i^0$.  Models \eqref{eq:over_soc_dyn} and \eqref{eq:under_soc_dyn} are completed by:
 \begin{align}
         0\leq p_{{\rm ch},i}^{k} + \overline{\Delta p}_{{\rm ch},i}^k \leq \overline{x}_{{\rm ch},i}^{k} P^{\rm max}_{{\rm ch},i} \qquad \forall i,k, \label{eq:bess_c4}\\
         \overline{x}_{{\rm dsc},i}^{k} P^{\rm min}_{{\rm dsc},i} \leq p_{{\rm dsc},i}^{k}+ \overline{\Delta p}_{{\rm dsc},i}^k \leq 0 \qquad\forall i,k, \label{eq:bess_c5}\\
         \overline{x}_{{\rm ch},i}^{k} + \overline{x}_{{\rm dsc},i}^{k} \leq 1 \qquad \forall i,k,  \label{eq:bess_c6}
\end{align}
and
\begin{align}
         0\leq p_{{\rm ch},i}^{k} + \underline{\Delta p}_{{\rm ch},i}^k \leq \underline{x}_{{\rm ch},i}^{k} P^{\rm max}_{{\rm ch},i} \qquad \forall i,k, \label{eq:bess_c7}\\
         \underline{x}_{{\rm dsc},i}^{k}P^{\rm min}_{{\rm dsc},i} \leq p_{{\rm dsc},i}^{k}+ \underline{\Delta p}_{{\rm dsc},i}^k \leq 0 \qquad\forall i,k, \label{eq:bess_c8}\\
         \underline{x}_{{\rm ch},i}^{k} + \underline{x}_{{\rm dsc},i}^{k} \leq 1 \qquad \forall i,k. \label{eq:bess_c9}
\end{align}
where $\overline{x}_{{\rm ch},i}^{k}$, $\overline{x}_{{\rm dsc},i}^{k}$, $\underline{x}_{{\rm ch},i}^{k}$, and $\underline{x}_{{\rm dsc},i}^{k}$ are binary variables that identifies the charge and discharge activation states for the two \ac{soc} trajectories.

It is easy to show that if
\begin{equation}\label{eq:bess_c10}
\overline{\soc}_{i}^k \leq \soc^{\rm max}_i, \qquad  \soc^{\rm min}_{i} \leq \underline{\soc}^k_i  \quad \forall i,k,
\end{equation}
then $\soc^{\rm min}_i \leq  \soc^k_i \leq \soc^{\rm max}_i$, where $\soc_i^{\rm max}$ and $\soc_i^{\rm min}$ are the desired limits for the battery \ac{soc}.

Finally, the following constraints are required to guarantee a maximum number of charging and discharging daily cycles equal to the prescribed parameters $\ell_i^{\rm ch}$ and $\ell_i^{\rm dsc}$, in order to limit the \ac{bess} ageing:
\begin{align}
         \eta_i^{\rm ch}\frac{\Delta t}{E_{{\rm b},i}^{\rm nom}} \sum_{k=0}^{T-1} \left( p_{{\rm ch},i}^{k} +\overline{\Delta p}_{{\rm ch},i}^k\right)&\leq \ell_i^{\rm ch} \quad \forall i, \label{eq:bess_c11} \\
                  -\eta_i^{\rm dsc}\frac{\Delta t}{E_{{\rm b},i}^{\rm nom}} \sum_{k=0}^{T-1} \left( p_{{\rm dsc}_i}^{k} +\underline{\Delta p}_{{\rm dsc},i}^k \right) &\leq \ell_i^{\rm dsc} \quad \forall i. \label{eq:bess_c12}
\end{align}

\subsection{TCLs model}\label{ssec:tcls_models}
Let us consider the $i$th \ac{tcl}, $i=1,2,\ldots,N_{\rm tcl}$. We suppose that this \ac{tcl} is working only as a cooling unit. However, with trivial changes, the proposed model can be used also for heating units. The basic variables, defined for any time step $k$ are: $\vartheta_{i}^k$ [\si{\celsius}], controlled temperature; $p_{{\rm tcl},i}^k$ [\si{\kilo\watt}], absorbed power by the \ac{tcl}; $x_{{\rm tcl},i}^k$, binary variable which identifies the activation state of the \ac{tcl}.

The time evolution of $\vartheta_i^k$ is modelled by:
\begin{equation}\label{eq:theta_dyn}
  \vartheta_{i}^k = \alpha_i \vartheta_{i}^{k-1} - \beta_i R_i \eta{^{\rm c}_i} p_{{\rm tcl },i}^{k-1} + \beta_i {\vartheta}_{{\rm ex},i}^{k-1}
\end{equation}
for $k=1,2,\ldots T-1$, given the initial condition $\vartheta_i^0.$ In \eqref{eq:theta_dyn}: $\eta{^{\rm c}_i}$ is the cooling efficiency, $R_i$ [\si{\celsius/(\kilo\watt)}] is the thermal resistance  $\alpha_i=\exp{-\Delta t /(C_i R_i)}$ and $\beta_i=1-\alpha_i$ are dynamic parameters, where $C_i$ [\si{\kWh /\celsius}] is the thermal capacitance of the controlled mass (air or water),
and ${\vartheta}_{{\rm ex},i}^{k}$ [\si{\celsius}] is the external temperature.
This last can be external from the \ac{lu} or be the temperature controlled by another \ac{tcl} within the \ac{lu}, \textit{i.e.}, $\hat{\vartheta}_{{\rm ex},i}^{k} = \vartheta_j^k$ for any $j\neq i$. In the first case, a day-ahead forecast profile $\{\hat{\vartheta}_{{\rm ex},i}^k\}_{k=0}^{T-1}$ is assumed to be available. Forecast errors are modelled by independent Gaussian random variables, \textit{i.e.},
\begin{equation}\label{eq:theta_ex_model}
  \vartheta_{{\rm ex},i}^{k} \sim \mathcal{N}(\hat{\vartheta}_{\rm ex}^k,\sigma_{{\rm ex},i}^k) \quad \forall i,k.
\end{equation}
It is worth remarking that model \eqref{eq:theta_ex_model} can be used also to represent the uncertainties of the dynamical model \eqref{eq:theta_dyn} and those introduced by other external disturbances such as windows and doors opening, people occupancy etc..

The model for the standard functioning of the considered \ac{tcl} is completed by the two following constraints:
\begin{equation}
    0 \leq p_{{\rm tcl},i}^k x_{{\rm tcl},i}^k\leq P^{\rm max}_{{\rm tcl},i}\qquad \forall i,k,
\end{equation}
which limits the power absorbed by the \ac{tcl} under the rated value ${P}^{\rm max}_{{\rm tcl},i}$; and
\begin{equation}\label{eq:tcl_c0}
    x_{{\rm tcl},i}^k \leq {\rm UP}_{{\rm tcl},i}^k \qquad \forall i,k,
\end{equation}
which assures that the \ac{tcl} is activated according to the user preferences, represented by the binary parameters ${\rm UP}_{{\rm tcl},i}^k$, defined as explained for the \acp{abp}.

Similarly to what done for \acp{bess} in Subsection~\ref{ssec:bess_model}, in order to quantify the potential energy reserve that can be provided by the \ac{tcl} during the day, the maximal positive and negative variations of the power consumption are introduced: $\overline{\Delta p}_{{\rm tcl},i}^k\geq 0$ and $\underline{\Delta p}_{{\rm tcl},i}^k\leq 0$ [\si{\kilo\watt}]. Given these two variations, it is possible to compute the time evolution of the over-bound and the under-bound temperatures $\overline{\vartheta}_{i}^k$ and $\underline{\vartheta}_{i}^k$ as it follows:
\begin{equation}\label{eq:tcl_c1}
\overline{\vartheta}_{i}^k = \alpha_i \overline{\vartheta}_{i}^{k-1} - \beta_i R_i \eta{^{\rm c}_i} \left( p_{{\rm tcl },i}^{k-1} + \underline{\Delta p}_{{\rm tcl},i}^{k-1}\right) + \beta_i \hat{\vartheta}_{{\rm ex},i}^{k-1}
\end{equation}
\begin{equation}\label{eq:tcl_c2}
\underline{\vartheta}_{i}^k = \alpha_i \underline{\vartheta}_{i}^{k-1} - \beta_i R_i \eta{^{\rm c}_i} \left( p_{{\rm tcl },i}^{k-1} + \overline{\Delta p}_{{\rm tcl},i}^{k-1}\right) + \beta_i \hat{\vartheta}_{{\rm ex},i}^{k-1}
\end{equation}
for all $i$ and $k=1,2,\ldots,T$, with initial conditions $\overline{\vartheta}_i^0$ and $\underline{\vartheta}_i^0$. The power variations must also satisfy the two following constraints:
\begin{align}
  0 \leq p_{{\rm tcl},i}^k +  \overline{\Delta p}_{\rm tcl}^k \leq x_{{\rm tcl},i}^k P^{\rm max}_{{\rm tcl},i}\qquad \forall i,k, \label{eq:tcl_c4}\\
  0 \leq p_{{\rm tcl},i}^k +  \underline{\Delta p}_{\rm tcl}^k \leq x_{{\rm tcl},i}^k P^{\rm max}_{{\rm tcl},i}\qquad \forall i,k. \label{eq:tcl_c5}
\end{align}

The main task required to a \ac{tcl} is to keep the controlled temperature within a desired comfort interval, which, for the $i$th \ac{tcl}, is indicated with $[\vartheta_i^{\rm min},\vartheta_i^{\rm max}]$. Because of the stochastic assumption made for the external temperature in \eqref{eq:theta_ex_model}, this objective can be assured only in probabilistic sense. In this paper, we use \textit{chance constraints} adopting the \textit{separation constraints approximation}~\cite{Conte:2018}. Thus, the \ac{tcl} objective is that, for all $k$,
\begin{align}
    \mathbf{P}\left( \vartheta_i^k \leq \vartheta_i^{\rm max} \right)\geq 1-r, \ \
    \mathbf{P}\left( \vartheta_i^k \geq \vartheta_i^{\rm min} \right)\geq 1-r, \label{eq:cc}
\end{align}
where $0<r<0.5$ is the so-called \textit{reliability}. Using \eqref{eq:tcl_c1}, \eqref{eq:tcl_c2}, and \eqref{eq:theta_ex_model}, chance constraints in \eqref{eq:cc} can be rewritten as the following deterministic constraints:
\begin{align}
    \overline{\vartheta}_i^k &\leq \vartheta_i^{\rm max}-\sigma_{{\rm ex},i}^{k-1}\sqrt{2}{\rm erf}^{-1}(1-2r)\quad \forall k, \label{eq:cc_1a} \\
    \underline{\vartheta}_i^k &\geq \vartheta_i^{\rm min}+\sigma_{{\rm ex},i}^{k-1}\sqrt{2}{\rm erf}^{-1}(1-2r)\quad \forall k, \label{eq:cc_2a}
\end{align}
where ${\rm erf}^{-1}(\cdot)$ is the inverse Gauss error function.

\subsection{RESs, UPDs, NCDs models}
The power generated by the $i$th \ac{res}, $i=1,2,\ldots,N_{\rm res}$, at time $k$, is indicated with $p_{{\rm res},i}^k$ [\si{\kilo\watt}]. A day-ahead forecast profile $\{ \hat{p}_{{\rm res},i}^k \}_{k=0}^{T-1}$ is supposed to be available.

The power consumed by the $i$th \ac{upd}, $i=1,2,\ldots,N_{\rm upd}$, at time $k$, is indicated with $p_{{\rm upd},i}^k$ [\si{\kilo\watt}]. For these devices, it is supposed that the user defines a fixed day plan $\{ \hat{p}_{{\rm upd},i}^k \}_{k=0}^{T-1}$. This assumption is reasonable for appliances like lights in industrial buildings or in the common spaces of commercial building.

The aggregated power consumed by the \acp{ncd} at time $k$ is indicated with $p_{\rm ncd}^k$ [\si{\kilo\watt}]. In order to consider the use of these devices in the \ac{lu}-\ac{dap} optimization problem, a forecast profile $\{ \hat{p}_{\rm ncd}^k \}_{k=0}^{T-1}$ is supposed to be available.

Forecast errors are modelled by independent Gaussian random variables. Therefore,
\begin{align}
    p_{{\rm res},i}^k &\sim \mathcal{N}(\hat{p}_{{\rm res},i}^k,\sigma_{{\rm res},i}^k) \quad \forall i,k, \label{eq:pres_forecast_model}\\
    p_{{\rm ncd}}^k &\sim \mathcal{N}(\hat{p}_{{\rm ncd}}^k,\sigma_{\rm ncd}^k) \quad \forall k. \label{eq:pncd_forecast_model}
\end{align}

\section{Load Unit Day-Ahead Planning} \label{sec:lu_dap}
Based on models developed in Section~\ref{sec:lu_models}, the total power expected to be exchanged by the $h$th \ac{lu} at any time step $k$ is given by:
\begin{equation}\label{eq:ph}
\begin{aligned}
{p}^{k}_{h} = & \sum_{i=1}^{N_{\rm abp}}\sum_{j=1}^{n_i}p_{ij}^k + \sum_{i=1}^{N_{\rm pev}}p_{i}^k +
 \sum_{i=i}^{N_{\rm bess}} \left ( p_{{\rm ch},i}^k + p_{{\rm dsc},i}^k \right)\\
&  + \sum_{i=1}^{N_{\rm tcl}} p_{{\rm tcl},i}^k -  \sum_{i=1}^{N_{\rm res}} \hat{p}_{{\rm res},i}^k +\sum_{i=1}^{N_{\rm upd}} \hat{p}_{{\rm upd},i}^k  + \hat{p}_{{\rm ncd},h}^k.
\end{aligned}
\end{equation}
As for energy, ${p}^{k}_{h}\geq 0$ means power import, and ${p}^{k}_{h}\leq 0$ means power export. Notice that in \eqref{eq:ph} powers $p_{ij}^k$, $p_i^k$, $p_{{\rm ch},i}^k$, $p_{{\rm dsc},i}^k$, $p_{{\rm tcl},i}^k$ are variables which has to be determined, whereas $\hat{p}_{{\rm res},i}^k$, $\hat{p}_{{\rm upd},i}^k$, and $\hat{p}_{{\rm ncd},h}^k$ are given forecast data.

Power $p_h^k$ has to be intended as the \textit{base} power expected to be exchanged by the \ac{lu}. Indeed, in Subsection~\ref{ssec:bess_model} and Subsection~\ref{ssec:tcls_models} the time-varying maximal positive and negative power variations $\overline{\Delta p}_{{\rm bess(tcl)},i}^k$, $\underline{\Delta p}_{{\rm bess(tcl)},i}^k$ have been introduced in order to dimension the potential provision of energy reserves from \acp{bess} and \acp{tcl}. Actually, these power variations are exploited for two different issues: 1) to effectively provide the total energy reserves $\overline{\Delta E}_h^k$ and $\underline{\Delta E}_h^k$ to be communicated to the \ac{agg}; and 2) to compensate the uncertainties introduced by the forecast errors of \acp{res} and \acp{ncd}. This is realized by setting, for all $i,k$,
\begin{align}
    \overline{\Delta p}_{{\rm bess(tcl)},i}^k &= \overline{\Delta p}_{{\rm bess(tcl),flx},i}^k + \overline{\Delta p}_{{\rm bess(tcl),unc},i}^k \\
    \underline{\Delta p}_{{\rm bess(tcl)},i}^k &= \underline{\Delta p}_{{\rm bess(tcl),flx},i}^k + \underline{\Delta p}_{{\rm bess(tcl),unc},i}^k
\end{align}
where the subscript `unc' means that the power variation is used to compensate the forecast errors, whereas the subscript `flx' means that the power variation is used to provide energy reserves to the \ac{agg}. Therefore, we can define, for the entire $h$th \ac{lu}, the power variations
\begin{align}
    \overline{\Delta p}_{{\rm unc},h}^k &= \sum_{i=1}^{N_{\rm bess}} \overline{\Delta p}_{{\rm bess,unc},i}^k + \sum_{i=1}^{N_{\rm tcl}} \overline{\Delta p}_{{\rm tcl,unc},i}^k \quad \forall k, \\
    \underline{\Delta p}_{{\rm unc},h}^k &= \sum_{i=1}^{N_{\rm bess}} \underline{\Delta p}_{{\rm bess,unc},i}^k + \sum_{i=1}^{N_{\rm tcl}} \underline{\Delta p}_{{\rm tcl,unc},i}^k \quad \forall k,
\end{align}
to be used for compensating the forecast errors. The maximal positive and negative energy reserves provided by the $h$th \ac{lu} are given by
\begin{equation}
    \overline{\Delta E}_h^k = \Delta t \cdot \overline{\Delta p}_{{\rm flx},h}^k, \quad \underline{\Delta E}_h^k = \Delta t \cdot \underline{\Delta p}_{{\rm flx},h}^k \quad \forall k,
\end{equation}
where
\begin{align}
    \overline{\Delta p}_{{\rm flx},h}^k &= \sum_{i=1}^{N_{\rm bess}} \overline{\Delta p}_{{\rm bess,flx},i}^k + \sum_{i=1}^{N_{\rm tcl}} \overline{\Delta p}_{{\rm tcl,flx},i}^k \ \ \forall k, \\
    \underline{\Delta p}_{{\rm flx},h}^k &= \sum_{i=1}^{N_{\rm bess}} \underline{\Delta p}_{{\rm bess,flx},i}^k + \sum_{i=1}^{N_{\rm tcl}} \underline{\Delta p}_{{\rm tcl,flx},i}^k  \ \ \forall k. \label{eq:lu_c1}
\end{align}

Based on \eqref{eq:pres_forecast_model} and \eqref{eq:pncd_forecast_model} the total forecast error for the power exchanged by the $h$th \ac{lu} at time $k$ results to be $\varepsilon_h^k \sim \mathcal{N}(0, \sigma_{{\rm unc},h}^k)$, where
$
    \sigma_{{\rm unc},h}^k = \sqrt{\sum_{i=1}^{N_{\rm res}} \sigma_{{\rm res},i}^k + \sigma_{\rm ncd}^k }.
$

In order to compensate this error, $\overline{\Delta p}_{{\rm unc},h}^k$ and $\underline{\Delta p}_{{\rm unc},h}^k$ are set to satisfy:
\begin{align}
    \mathbf{P} \left( \overline{\Delta p}_{{\rm unc}}^k \geq \varepsilon_h \right) &\geq 1-r \quad \forall k, \\
    \mathbf{P} \left( \underline{\Delta p}_{{\rm unc}}^k \leq -\varepsilon_h \right) &\geq 1-r  \quad \forall k,
\end{align}
which correspond to
\begin{align}
   \overline{\Delta p}_{{\rm unc},h}^k &\geq \sigma_{{\rm unc},h}^k \sqrt{2}{\rm erf}^{-1}(1-2r)  \quad \forall k,  \label{eq:lu_c2}\\
   \underline{\Delta p}_{{\rm unc},h}^k &\leq -\sigma_{{\rm unc},h}^k \sqrt{2}{\rm erf}^{-1}(1-2r)  \quad \forall k.
\end{align}

The base power exchange and the introduced power variations defined for the $h$th \ac{lu} must also satisfy the minimum and maximum power limits $P_h^{\rm min}\leq 0$ and $P_h^{\rm max}\geq 0$ [\si{\kilo\watt}]:
\begin{align}
     p_h^k + \overline{\Delta p}_{{\rm flx},h} + \overline{\Delta p}_{{\rm unc},h} &\leq  P_h^{\rm max}, \quad \forall k, \\
      p_h^k + \underline{\Delta p}_{{\rm flx},h} + \underline{\Delta p}_{{\rm unc},h} &\geq  P_h^{\rm min}, \quad \forall k.
\end{align}
Notice that $P_h^{\rm min}$ is supposed to be negative or null. When it is not null, it means that power export is allowed.

Finally, since the prices paid for importing energy and received for exporting energy are generally different, the base power exchange is partitioned as
\begin{equation}
    p_h^k = p_{{\rm imp},h}^k + p_{{\rm exp},h}^k \quad \forall k,
\end{equation}
where, by introducing the binary variable $x_{\rm imp}^k$,
\begin{align}
  p_{{\rm imp},h}^k &\leq x_{\rm imp}^k P_h^{\rm max} \quad \forall  k, \\
  p_{{\rm exp},h}^k &\geq (1-x_{\rm imp}^k) P_h^{\rm min} \quad \forall  k;
\end{align}
therefore, the resulting energies imported end exported within the $k$th sampling interval are:
\begin{equation}\label{eq:lu_c3}
    E_{{\rm imp},h}^k = \Delta t \cdot  p_{{\rm imp},h}^k, \quad  E_{{\rm exp},h}^k = \Delta t \cdot  p_{{\rm exp},h}^k \quad \forall k.
\end{equation}

\subsection{LU-DAP optimization problem}
We have now all the elements for providing the \ac{lu}-\ac{dap} optimization problem for the $h$th \ac{lu}, which consists in the \textit{minimization} of the cost function:
\begin{equation}
    J_h = \sum_{k=0}^{T-1} c_{{\rm imp},h}^k E_{{\rm imp},h}^k - c_{{\rm exp},h}^k E_{{\rm exp},h}^k -c_{\rm flx}\tensor*[^k]{c}{*_{\rm flx}}\left( \overline{\Delta E}_{h}^k - \underline{\Delta E}_{h}^k \right) \nonumber
\end{equation}
such that the following constraints are satisfied: \eqref{eq:abp_c1}--\eqref{eq:pev_c3}, \eqref{eq:bess_c1}--\eqref{eq:bess_c12}, \eqref{eq:tcl_c0}--\eqref{eq:tcl_c5}, \eqref{eq:cc_1a}--\eqref{eq:cc_2a}, \eqref{eq:ph}--\eqref{eq:lu_c1}, \eqref{eq:lu_c2}--\eqref{eq:lu_c3}.

This optimization problem is linear mixed-integer. The result of the optimization are the power profiles of all the controllable devices, and the maximal positive and negative power variations of \acp{bess} and \acp{tcl}, together with the related aggregated values for the $h$th \ac{lu}. In the paper, the optimized variables are indicated with $\widehat{(\cdot)}$.

If energy export is not enabled, with a fixed price $\tensor*[^k]{c}{*_{\rm flx}}$ paid both for positive and negative energy reserves, the \ac{lu} will choose not to provide any negative reserve to minimize the cost function. Therefore, in this cases, if a minimum amount of negative energy reserve is required, an additive constraint should be added. For example, positive and negative reserves could be forced to be equal:
\begin{equation} \label{eq:equal_reserve}
    \overline{\Delta E}_h^k=\underline{\Delta E}_h^k, \quad \forall k.
\end{equation}

\subsection{Intra-Day operation}
As illustrated in Section~\ref{sec:aggregate_dap_and_day_operation},
once the \ac{lu}-\ac{dap} optimization problem has been executed, the $h$th \ac{lu} sends to the \ac{agg} the resulting optimized energy exchange $\widehat{E}_h^k= \widehat{E}_{{\rm imp},h}^k +  \widehat{E}_{{\rm exp},h}^k$ and the optimized energy reserves $\widehat{\overline{\Delta E}}_h^k$ and $\widehat{\underline{\Delta E}}_h^k$.

During the considered day, the \ac{agg} sends to the $h$th \ac{lu} the energy reserve reference signal $\Delta E_{{\rm ref},h}^k$, defined in \eqref{eq:Eref}.
At the same time, the power consumption of \acp{res} and \acp{ncd} is monitored and the forecasts error $\varepsilon_h^k$ is computed.
\acp{bess} and \acp{tcl} are therefore called to provide the required energy reserves and to compensate the forecasts error proportionally to the percentage of their own (optimized) maximum power variations.

\section{Simulation Results} \label{sec:simulation_results}
\begin{figure}[htbp!]
     \centering
     \includegraphics[trim={1.7cm 2cm 2.8cm
0.2cm},clip=true,width=1\linewidth]{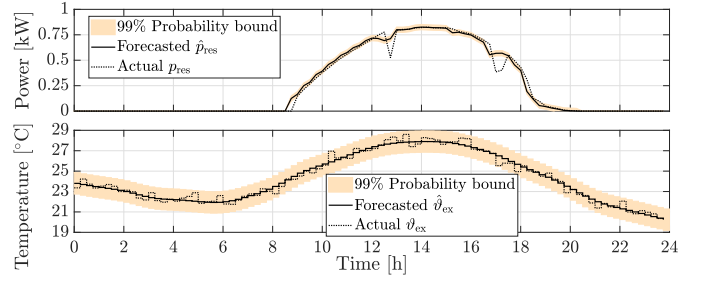}
     \caption{Forecasts of \ac{pv} generation (top) and external temperature (bottom).}
     \label{fig:PV_Temp}
\end{figure}

To analyze the effectiveness of the proposed approach, the case of an aggregate of $H$=200 houses has been considered.
Simulation parameters are reported in Table~\ref{table:table_parms}. Each house (\textit{i.e.}, each \ac{lu}) is equipped with two \acp{abp} (a dish washer and a washing machine), a \ac{bess}, an inverter-driven air cooling system (\ac{tcl}), a \ac{pv} generator (\ac{res}), and  a \ac{pev}. Table~\ref{table:table_parms} reports the parameters common to all the houses, which are then differentiated by randomly generating the initial \acp{bess}' \acp{soc}, the initial internal air temperatures, the temperature set-points, the amount of required \ac{pev} recharging ($\Delta \rm{SoC}^{\rm p}_1$), and the time preferences for the \ac{pev} recharge. All \acp{tcl} are switched off from 0\textsc{am} to 8\textsc{am} and from 8\textsc{pm} to 12\textsc{pm}. \acp{res} have the same rated power and are supposed to have the same generation profile, depicted in Fig.~\ref{fig:PV_Temp}. This figure also reports the \ac{pv} generation and external temperature forecasts used by the \ac{lu}-\ac{dap} optimization.

Notice that the power export is not enabled ($P^{\rm min}_h=0, \forall h$). Moreover, consider that the energy prices are constant (\textit{i.e.}, $\tensor*[^k]{c}{*_{\rm flx}} = c_{\rm flx}$) and the price paid by the \ac{agg} for the energy reserve is five times the cost paid for the imported energy (\textit{i.e.}, $\tensor*[^k]{c}{*_{\rm flx}^{\rm agt}} = 5c_{\rm flx}$). Thus, the \acp{lu} have a potential significant economical advantage in providing the energy reserve. The values of these prices have been derived from the Italian Regulation Authority (ARERA). In order to obtain a negative energy reserve, the additional constraint \eqref{eq:equal_reserve} has been included in the optimization problem.

Simulation have been implemented on the MATLAB platform. The \ac{lu}-\ac{dap} opimization problem has been written using the AMPL language and solved with the IBM ILOG Cplex solver.

\begin{figure}[htbp!]
     \centering
     \includegraphics[trim={1.9cm 0.1cm 2.8cm
0.cm},clip=true,width=1\linewidth]{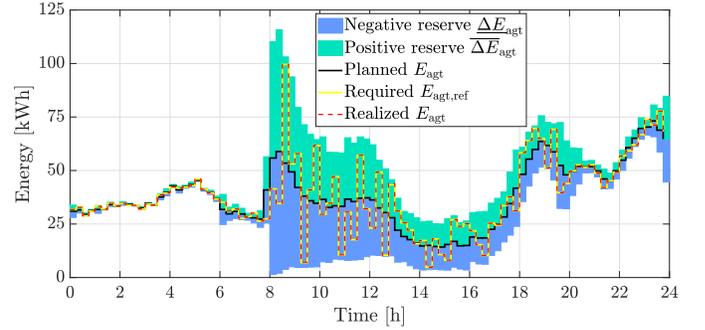}
     \caption{Resulting day-ahead plan and \ac{dr} signal realization for the \ac{agg}.}
     \label{fig:agg_flex}
\end{figure}

Figure~\ref{fig:agg_flex} reports the results obtained for the \ac{agg}. It appears clear that until the \acp{tcl} are switched on at 8\textsc{pm}, the offered energy reserves are very small, whereas, during the daylight hours, when also \acp{res} are active, the \ac{agg} is able to offer a significant amount of energy reserve. Figure \ref{fig:agg_flex} also reports an example of a possible required \ac{dr} signal perfectly realized by the \ac{agg}.

\begin{figure}[htbp!]
     \centering
     \includegraphics[trim={1.7cm 0.1cm 2.8cm
0.2cm},clip=true,width=1\linewidth]{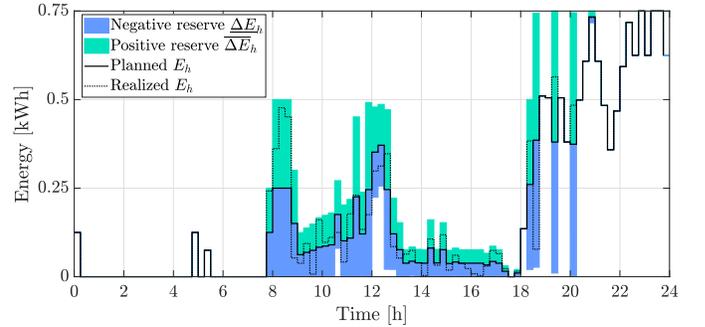}
     \caption{Energy profiles for one of the 200 houses.}
     \label{fig:home_flex}
\end{figure}

\begin{figure}[htbp!]
     \centering
     \includegraphics[trim={1.7cm 0cm 2.8cm
0cm},clip=true,width=1\linewidth]{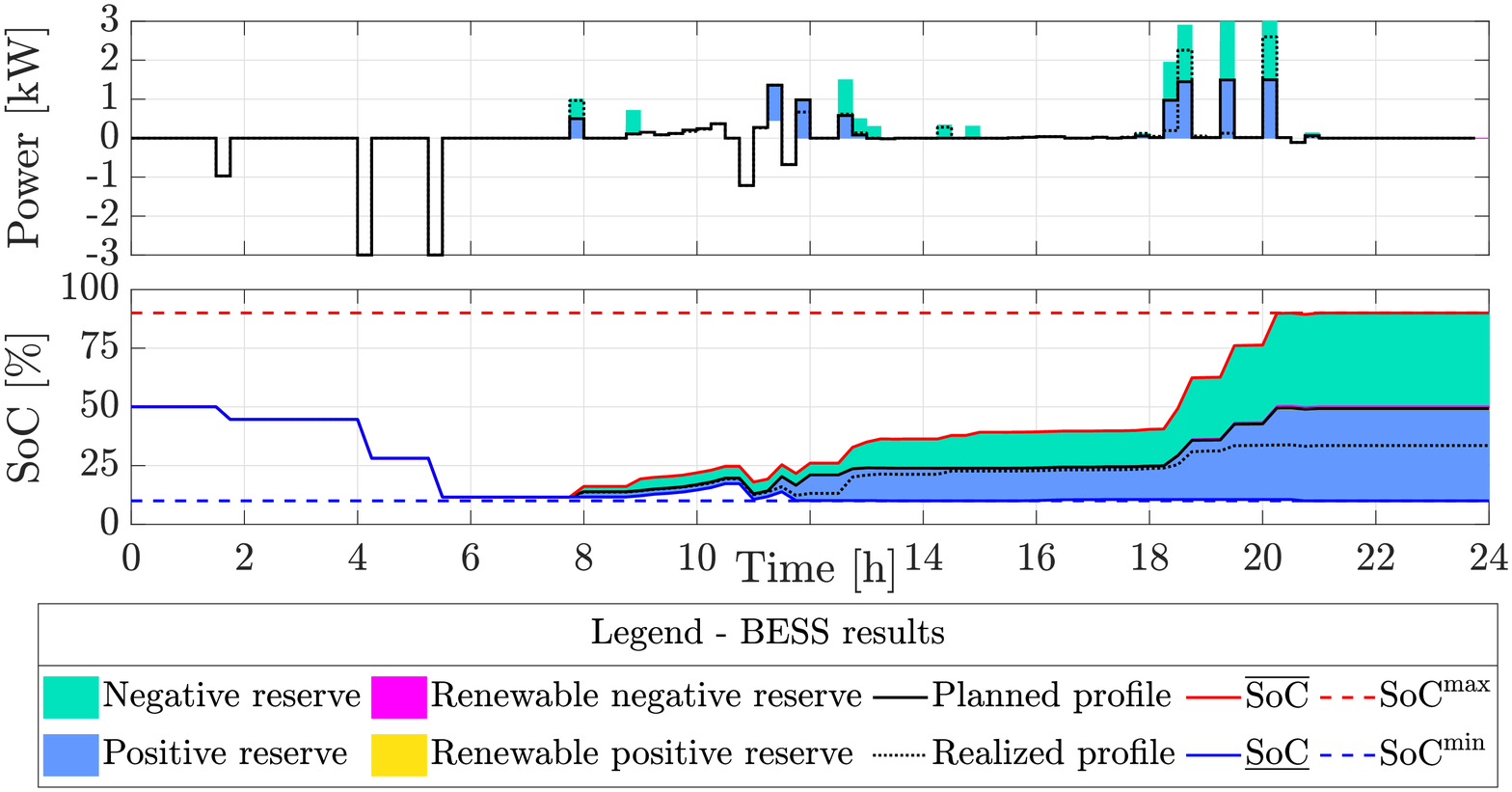}
     \caption{\ac{bess} results for one of the 200 houses.}
     \label{fig:BESS}
\end{figure}

\begin{figure}[htbp!]
     \centering
     \includegraphics[trim={1.4cm 0cm 2.8cm
0cm},clip=true,width=1\linewidth]{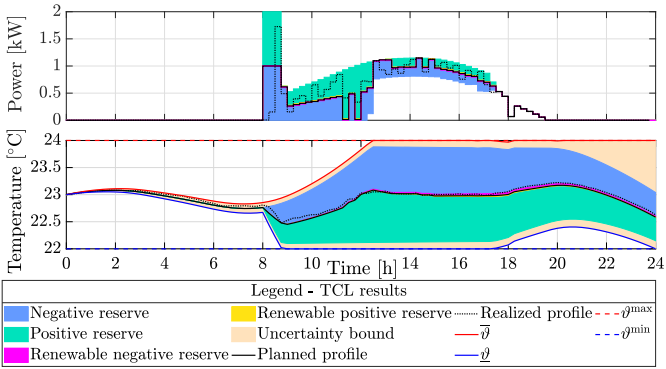}
     \caption{\ac{tcl} results for one of the 200 houses.}
     \label{fig:TCL}
\end{figure}

Figures \ref{fig:home_flex}--\ref{fig:TCL} report the results obtained for one of the 200 houses. In particular, Fig.~\ref{fig:home_flex} shows the planned and realized energy profiles and the offered energy reserves. Figure~\ref{fig:BESS} reports the corresponding \ac{bess} power exchange and \ac{soc} trajectories, whereas Fig.~\ref{fig:TCL} depicts the ones of the \ac{tcl} power consumption and of the internal temperature. In these two figures, the contribution of the \ac{bess} and of the \ac{tcl} to the energy reserves and the potential consequences on the battery \ac{soc} and on the internal air temperature are also shown. It is worth noting that, for the considered house, the \ac{pev} recharge is active during the first hours and at the end of the day (the \ac{pev} recharge profile is not reported for the sake of brevity). This leads to discharge the \ac{bess} in the early morning, without providing significant energy reserves. Then, during the daylight hours, the \ac{tcl} manages the thermal energy to provide a significant amount of energy reserve. Finally, in Fig.~\ref{fig:PV_Temp}(bottom) it is shown how the temperature forecast uncertainty is taken into account in the definition temperature profiles.

\begin{table}[t]
  \centering
  \tiny{
\caption{Simulation parameters.}
\begin{tabular}{l c c}
\hline
Description & Symbol &  Value \\
\hline\vspace{-6pt}\\
 \multicolumn{3}{l}{\textbf{ABP}} \\
ABP number & $N_{\rm{abp}}$ & 2 \\
Phases number & $n_i$ & [4,4]\\
Energy used by each phase [\si{\kWh}] & $E_{ij}$ & $ \begin{bmatrix}
0.11 & 0.2 & 0.07 & 0.8 \\
0.11 & 0.2 & 0.07 & 0.8
\end{bmatrix} $\\
Intervals for each phase & $\overline{T}_{ij}$ & $ \begin{bmatrix}
3 & 1 & 2 & 2 \\
3 & 1  & 2 & 2
\end{bmatrix} $\\
Max. power for each phase [\si{\kilo\watt}] & ${P}^{\rm max}_{ij}$ & $ \begin{bmatrix}
0.15 & 1.6 & 0.15 & 1.6 \\
0.15 & 1.6 & 0.15 & 1.6
\end{bmatrix} $ \\
Min. power for each phase [\si{\kilo\watt}] & ${P}^{\rm min}_{ij}$ & $0 \ \forall i,j$ \\
 \multicolumn{3}{l}{\textbf{PEV}} \\
\vspace{-6pt}\\
PEV number & $N_{\rm{pev}}$ & 1 \\
Nominal energy [\si{\kWh}] & $E_{\rm{p},1}^{\rm{nom}}$ & 15\\
Recharge efficiency [p.u.] & $\eta_1$ & 0.9\\
Max. recharge power [\si{\kilo\watt}] & ${P}^{\rm max}_1$ & 3.3\\
\vspace{-6pt}\\
 \multicolumn{3}{l}{\textbf{BESS}} \\
 \vspace{-6pt}\\
 BESS number & $N_{\rm{bess}}$ & 1 \\
Nominal energy [\si{\kWh}] & $E_{\rm{b},1}^{\rm{nom}}$ & 5\\
Recharge efficiency [p.u.] & $\eta_1^{\rm{ch}}$ & 0.9\\
Discharge efficiency [p.u.] & $\eta_1^{\rm{dsc}}$ & 1.1\\
Max. recharge power [\si{\kilo\watt}] & ${P}^{\rm max}_{\rm{ch},1}$ & 3\\
Max. discharge power [\si{\kilo\watt}] & ${P}^{\rm min}_{\rm{dsc},1}$ & -3\\
Min. SoC [\%] & ${\rm{SoC}}_{1}^{\rm{max}}$ & 90\\
Max. SoC [\%] & ${\rm{SoC}}_{1}^{\rm{min}}$  & 10\\
Max. number of charging cycle  & $\ell_1^{\rm{ch}}$ & 1\\
Max. number of discharging cycle  & $\ell_1^{\rm{dsc}}$ & 1\\
\vspace{-6pt}\\
 \multicolumn{3}{l}{\textbf{TCL}} \\
\vspace{-6pt}\\
TCL number & $N_{\rm{tcl}}$ & 1 \\
Thermal resistance [\si{\celsius/(\kilo\watt)}] & $R_1$ & 2.5$\times 10^{-6}$
\\
Thermal capacitance [\si{\kilo \watt \second /\celsius}] & $C_1$ & 14400 \\
Cooling efficiency [p.u.] & $\eta_1^{\rm c}$ & 2 \\
Rated power [\si{\kilo\watt}] & ${P}^{\rm max}_{{\rm{tcl}},1}$ & 2 \\
Temperature forecast std. dev. & $\sigma_{\rm{ex},1}$ & 0.1 \\
\vspace{-6pt}\\
 \multicolumn{3}{l}{\textbf{LU}} \\
\vspace{-6pt}\\
Max. imported power [\si{\kilo\watt}] & $P_h^{\rm{max}}$  & 3\\
Max. exported power [\si{\kilo\watt}] & $P_h^{\rm{min}}$  & 0\\
PV rated power [\si{\kilo\watt}] & - & 1\\
Chance-constraint reliability [p.u.] & $r$ & 0.05\\
\vspace{-6pt}\\
 \multicolumn{3}{l}{\textbf{DAP}} \\
\vspace{-6pt}\\
LU number & $H$ & 200 \\
Time horizon [h] & $T$ & 24 \\
Sampling time [h] &$\Delta t$ & 0.25 \\
Imported energy price [\EUR{}/\si{\kWh}] & $c_{h,{\rm{imp}}}$ & 0.2\\
AGT reserve price [\EUR{}/\si{\kWh}] & $c_{\rm{flx}}^{\rm{agt}}$ & 30\\
LU reserve price [\EUR{}/\si{\kWh}] & $c_{\rm{flx}}$ & 1\\

\hline
\end{tabular}
 \label{table:table_parms}
 }
\end{table}

%
\section{Conclusion}\label{sec:conclusions}
The paper describes a decentralized optimization framework for demand response of the aggregation of load units. The strength of the proposed approach lies in the fact that the optimization is performed by each end user, assuming a complete knowledge of equipment and desires. The outcome of the process is a desired profile, and a reserve, which is, by construction, compliant with all the constraints the end users have designed for their appliances. This reserve can be used not only upon request, but also to compensate modeling and forecast errors.

\bibliographystyle{IEEEtran}

\end{document}